\documentclass[prb,aps,nopacs,superscriptaddress,onecolumn]{revtex4}
\usepackage{color} 
\usepackage{times}
\usepackage{amssymb} 
\usepackage{amsmath} 
\usepackage{ragged2e}
\usepackage{graphicx}
\usepackage{epstopdf}
\usepackage[english]{babel}
\usepackage{mathrsfs}
\usepackage{textcomp}
\usepackage{amsthm}
\usepackage{amsfonts}
\usepackage{soul}
\usepackage{braket}

\newcommand{\be}{\begin{equation}}
\newcommand{\ee}{\end{equation}}
\newcommand{\bea}{\begin{eqnarray}}
\newcommand{\eea}{\end{eqnarray}}
\newcommand{\ba}{\begin{array}}
\newcommand{\ea}{\end{array}}
\newcommand{\lf}{\lfloor}
\newcommand{\rf}{\rfloor}
\newcommand{\change}[1]{\textcolor{black}{#1}}

\begin{document}
\title{Dynamics and correlations in Motzkin and Fredkin spin chains}

\author{L. Dell'Anna}
\affiliation{Dipartimento di Fisica e Astronomia ``G. Galilei'',
Universit\`a di Padova, via F. Marzolo 8, I-35131, Padova, Italy}

\author{L. Barbiero}
\affiliation{Center for Nonlinear Phenomena 
and Complex Systems, 
Universit\'e Libre de Bruxelles, CP 231, 
Campus Plaine, B-1050 Brussels, Belgium}

\author{A. Trombettoni}
\affiliation{CNR-IOM DEMOCRITOS Simulation Center, Via Bonomea 265, I-34136,Trieste,\\ 
SISSA and INFN, Sezione di Trieste, Via Bonomea 265, I-34136 Trieste, Italy}

\begin{abstract}
The Motzkin and Fredkin quantum spin chains are described by frustration-free Hamiltonians recently introduced and studied because of their anomalous behaviors in the correlation functions and in the entanglement properties.   
In this paper we analyze their quantum dynamical properties, focusing in particular on the time evolution of the excitations driven by a quantum quench,   looking at the correlations functions of spin operators defined along different directions, and discussing the results in relation with the cluster decomposition property.
\end{abstract}

\maketitle

\section{Introduction}

The study of quantum systems driven out of equilibrium has attracted a lot of attention in the last few years. 
Theoretical and experimental interest on how fast the correlations can spread in quantum many-body systems  has been renewed \cite{polkonikov11,eisert15,huse15,kollath17} after the proof, for critical theories, that the maximum velocity of the spreading of correlations is given by the group velocity in the final gapless system \cite{pasquale06}.  
Actually, the existence of a maximal velocity known as the Lieb-Robinson bound \cite{lieb}, has been shown to exist theoretically in several interacting many-body systems, due to short-range interactions which limit the propagation of information making finite its spreading speed. Once the system is subject to a sudden change of the parameters of a short-range Hamiltonian, often denoted as a quantum quench, the time evolution of two-point correlation functions shows a well defined light-cone-like propagation defining causally connected regions, up to exponentially small deviations.\\
This behavior is tightly related to the concept of locality which plays a crucial role in physical theories, with far reaching consequences, the most fundamental being the cluster decomposition property (CDP) \cite{hastings,nachtergaele}. The CDP implies that two-point connected correlations functions go to zero when the separation of the points goes to infinity. This is the reason why two parts of a system very far apart, separated by a large distance, behave independently. This property is expected to be verified for non-degenerate states in systems described by local Hamiltonians. \\
On the other hand, the presence of long-range interaction may cause 
the violation of the Lieb-Robinson bound and the presence of power-law tails outside the light-cone 
\cite{eisert13}. 
For these reasons the study of non-local properties 
and their effects on the light-cone in the time evolution driven by a quantum quench are certainly at the present date a very 
interesting field of research. \\
Recently novel quantum spin models have been introduced \cite{bravyi12, ramis16, luca16, olof17}. These models, referred to as Motzkin and Fredkin models, in spite of being described by local Hamiltonians and admitting a unique ground state, may exhibit violation of CDP and of the area law for the entanglement entropy, with the presence of anomalous and extremely fast propagation of the excitations after driving the system out-of-equilibrium \cite{luca16}. 
Very recently, also modified versions of these models have been introduced and studied \cite{ramis17,sugino}, 
which can exhibit a quantum phase transition separating an extensively entangled phase \cite{klich,olof2} from a topological one \cite{lucaB}. 
It has been given also a continuum description for the ground-state wavefunctions of those models, which can reproduce quite well some quantities as for example the local magnetization and the entanglement entropy, and whose scaling Hamiltonian is not conformally invariant \cite{fradkin}. Entanglement properties in these models have also been extensively studied, particularly the Renyi entropy \cite{sugino18}, the negativity and the mutual information, revealing a large-distance entanglement behavior \cite{luca19}, which can produce intriguing out-of equilibrium properties \cite{barbiero18}.

In a previous work \cite{luca16} we showed that, looking at the connected correlation functions of spins along $z$-directions ($S_Z$-$S_Z$ correlators) the ground states exhibit a violation of the cluster decomposition for the Motzkin and the Fredkin models with spins higher than $1$ (the so-called colorful versions of these models). This behavior goes with a square-root violation of the area law for the entanglement entropy and an extremely fast spreading of the excitations after a quantum quench.  For spin $1/2$ in the Fredkin model and spin $1$ in the Motzkin model (the so-called colorless cases), instead, the $S_Z$-$S_Z$ connected correlation functions vanish for long distances and their dynamics is characterized by a clear light-cone-like evolution after switching a local perturbation along the $z$-direction. 
The question we address in this paper is whether one can have absence of cone and extremely fast excitation spreading in the colorless
  cases. What we will show is that, in the same situation where there is a conventional light-cone dynamics for the $S_Z$-$S_Z$ correlators,  
one can still observe anomalous dynamics looking at the transverse spin correlations, e.g. the time evolution of the $S_X$-$S_X$ correlation functions, which also at equilibrium does not fulfill the cluster decomposition property. 
This analysis shows that also in the colorless version of the Motzkin and Fredkin models (for spin $1$ and $1/2$, respectively) there is a violation of the cluster decomposition along the $x$-direction and the absence of a light-cone behavior of the propagation of excitations looking at the correlations of the spins in the transverse directions. This result is in agreement with what found for the mutual information, which, at least for conformal theories, represents an upper bound for any connected correlation functions \cite{pasquale11}, and which in our cases remains finite even when measured between infinitely distant separated regions \cite{luca19}. 

\section{The models}
In this section we will briefly review the spin models under study, considering only the case with spin $1$ for the Motzkin model and and spin $1/2$ for the Fredkin model. In both cases the full Hamiltonian can be written as
\be
H=H_0+H_\partial
\ee
where $H_0$ is the bulk Hamiltonian and $H_\partial$ is the boundary term which remove the huge degeneracy of the ground state of $H_0$, making it uniquely defined. 

\subsection{Motzkin model}
The Motzkin Hamiltonian
\cite{bravyi12,ramis16} can be written as
a local Hamiltonian made by a bulk contribution, 
\bea
&& H_0=\frac{1}{2}
\sum_{j=1}^{L-1}
\left\{{\sf P}
\Big(\Ket{0_j \Uparrow_{j+1}} -\Ket{\Uparrow_{j} 0_{j+1}}\Big)
+{\sf P}\Big(\Ket{0_j \Downarrow_{j+1}} -\Ket{\Downarrow_{j} 0_{j+1}}\Big)
+{\sf P}\Big(\Ket{0_j 0_{j+1}} -\Ket{\Uparrow_{j} \Downarrow_{j+1}}\Big)\right\}\\
&&H_\partial={\sf P}\left(\Ket{\Downarrow_1}\right)+{\sf P}\left(\Ket{\Uparrow_L}\right),
\eea
where ${\sf P}(|.\rangle)$ denotes the projector $|.\rangle\langle.|$,
and $\Ket{\Uparrow}$ ($\Ket{\Downarrow}$) the $1$-spin up (down). $L$ is the size of the chain, namely the total number of sites. 
The  Hamiltonian is constructed in terms of projection operators which commute with the total spin $S^{tot}_Z=\sum_{j=1}^L S_Z(j)$ where
$S_Z(j)\Ket{\Uparrow_j} =\Ket{\Uparrow_j}$, 
$S_Z(j)\Ket{\Uparrow_j} =0$, and 
$S_Z(j)\Ket{\Downarrow_j} =-\Ket{\Downarrow_j}$. 
The unique ground-state, $\ket{M}$, is known exactly and corresponds to an equal-weight superposition of states defined through Motzkin paths \cite{bravyi12,ramis16}, so that $S^{tot}_Z\ket{M}=0$, namely it belongs to the zero-total spin sector. 
These states are such that, denoting the spins up, ${\Uparrow}$, by {/}, the spins down, ${\Downarrow}$, by ${\backslash}$ and spins zero, ${0}$, by $-$, one can construct a Motzkin path. A Motzkin path is any path on a $x$-$y$ plan connecting the origin $(0, 0)$ to the point $(0, L)$ with steps $(1, 0)$, $(1, 1)$, $(1, -1)$, where $L$ is an integer number. Any point $(x,y)$ of the path is such that $x$ and $y$ are not negative.\\

\subsection{Fredkin model}
For spin $1/2$ we will consider the following Fredkin model \cite{luca16,olof17}
\bea
\label{fred0}
&&H_0=\frac{1}{2}
\sum_{j=1}^{L-2}\left[
{\sf P}
\Big(\Ket{\downarrow_j \uparrow_{j+1} \downarrow_{j+2}} -
\Ket{ \uparrow_{j} \downarrow_{j+1}\downarrow_{j+2}} \Big)
+{\sf P}
\Big(\Ket{\uparrow_j \uparrow_{j+1} \downarrow_{j+2}} -
\Ket{\uparrow_{j} \downarrow_{j+1}\uparrow_{j+2} } \Big)
\right]\\
&&H_\partial={\sf P}\left(\Ket{\downarrow_1}\right)+{\sf P}\left(\Ket{\uparrow_L}\right).
\eea
Also in this case the Hamiltonian is constructed in terms of projection operators which commute with the total spin $S^{tot}_{Z}=\sum_{j=1}^L S_Z(j)$ where
$S_Z(j)\Ket{\uparrow_j} =\frac{1}{2}\Ket{\uparrow_j}$ and 
$S_Z(j)\Ket{\downarrow_j} =-\frac{1}{2}\Ket{\downarrow_j}$. 
The unique ground-state, $\ket{D}$, is known exactly and corresponds to an equal-weight superposition of states defined through Dyck paths \cite{luca16,olof17}, such that it belongs to the zero-total spin sector, $S^{tot}_Z\ket{D}=0$. 
These states are such that, denoting the spins up, ${\uparrow}$, by {/} and the spins down, ${\downarrow}$, by ${\backslash}$, one can construct a Dyck  path. A Dick path is any path on a $x$-$y$ plan connecting the origin $(0, 0)$ to the point $(0, L)$ with steps $(1, 1)$, $(1, -1)$, where $L$ is an even integer number. Any point $(x,y)$ of the path is such that $x$ and $y$ are not negative.\\

\section{Dynamics after a quantum quench}
\begin{figure}[!ht]
\includegraphics[width=6.cm]{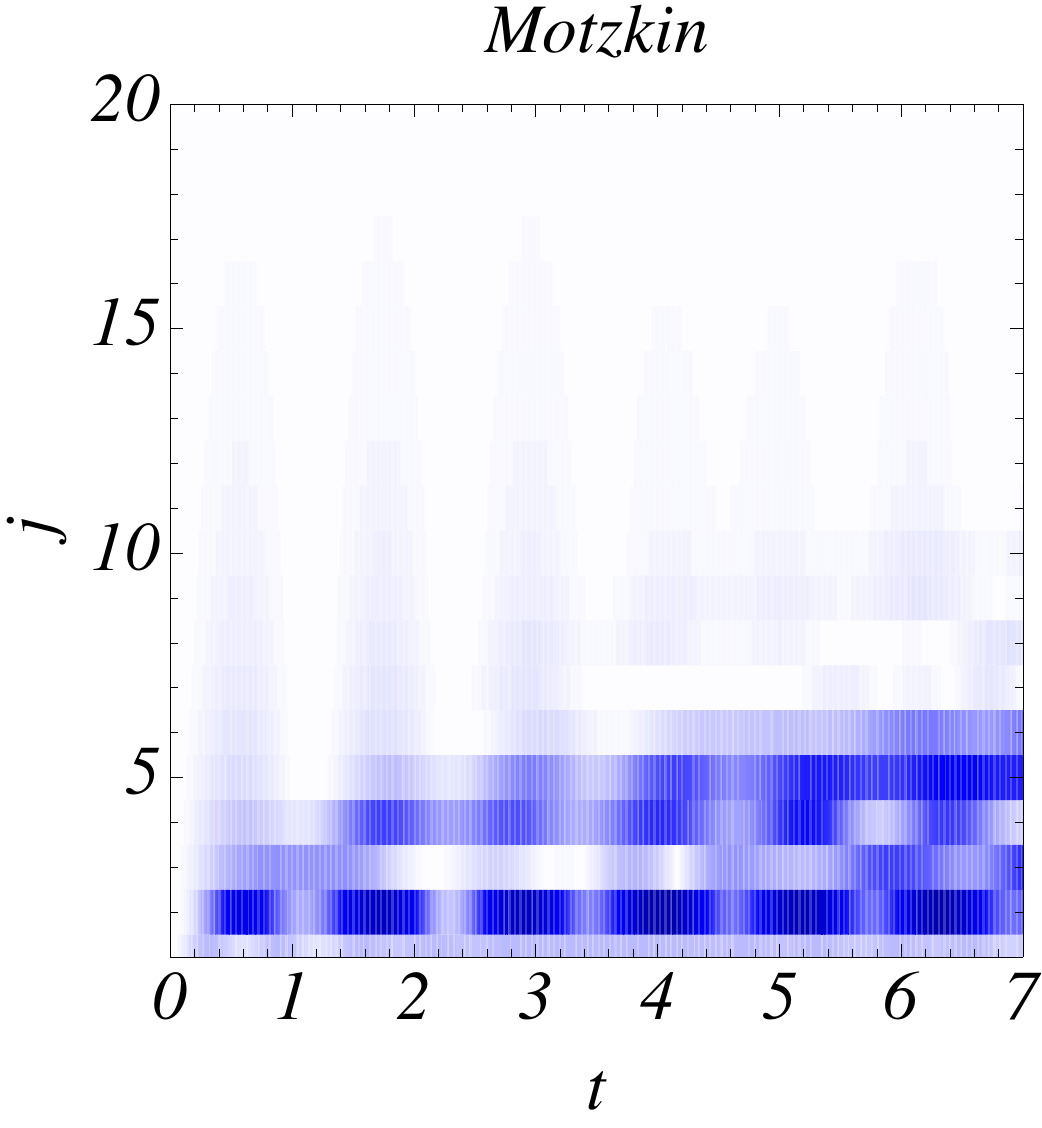}\hspace{1cm}
\includegraphics[width=6.cm]{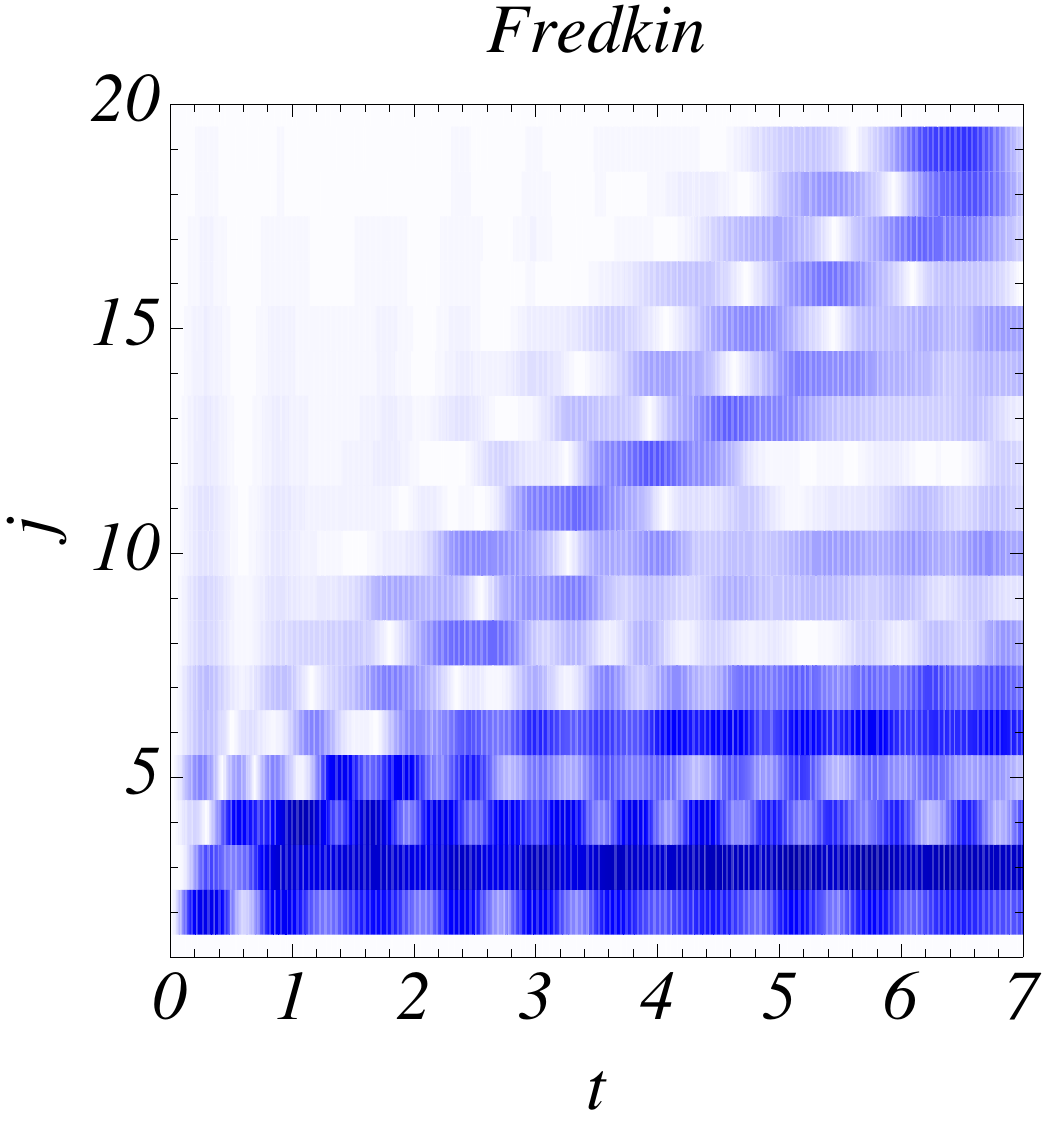}
\caption{
(Left) Time evolution of $\big|\langle S_Z(1,t) S_Z(j,t)\rangle_c - \langle S_Z(1,0) S_Z(j,0) \rangle_c\big|$
after switching on, at time $t=0$, the local field $h_z \,S_Z(j_0)$ on the site $j_0=1$ of the Motzkin spin chain ($L=20$, $h_z=5$).
(Right) Time evolution of $\big|\langle S_Z(2,t) S_Z(j,t)\rangle_{c} - \langle S_Z(2,0) S_Z(j,0)\rangle_{c}\big|$ 
after switching on, at time $t=0$, the local field $h_z \,S_Z(j_0)$ on the site 
$j_0=2$ of the Fredkin spin chain ($L=20$, $h_z=5$). The colors are chosen so that white correspond 
to zero and blue of increasing intensity to increasing values. 
}
\label{figure_zz}
\end{figure}

\begin{figure}[!ht]
\includegraphics[width=6.cm]{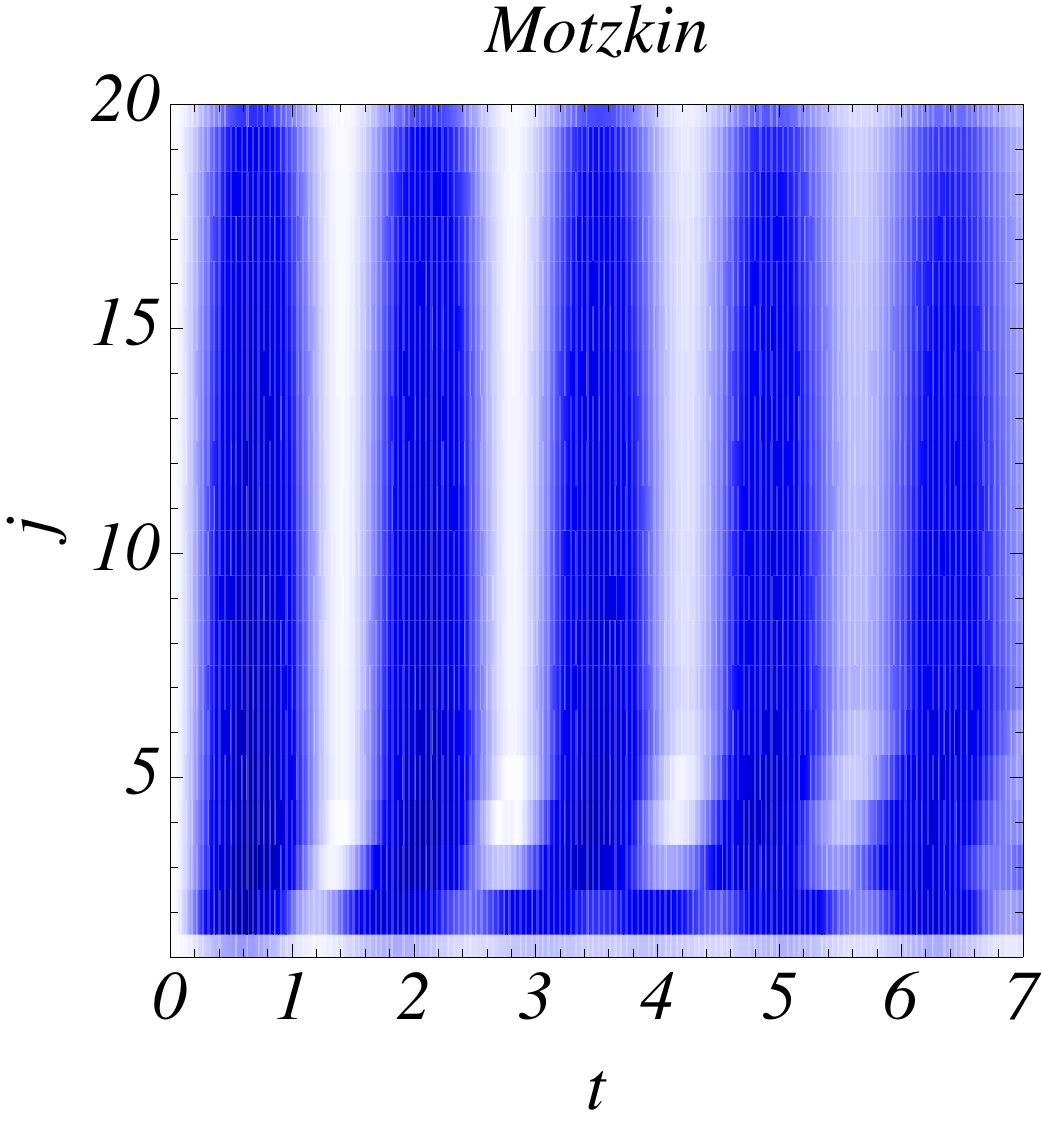}
\hspace{1cm}
\includegraphics[width=6.cm]{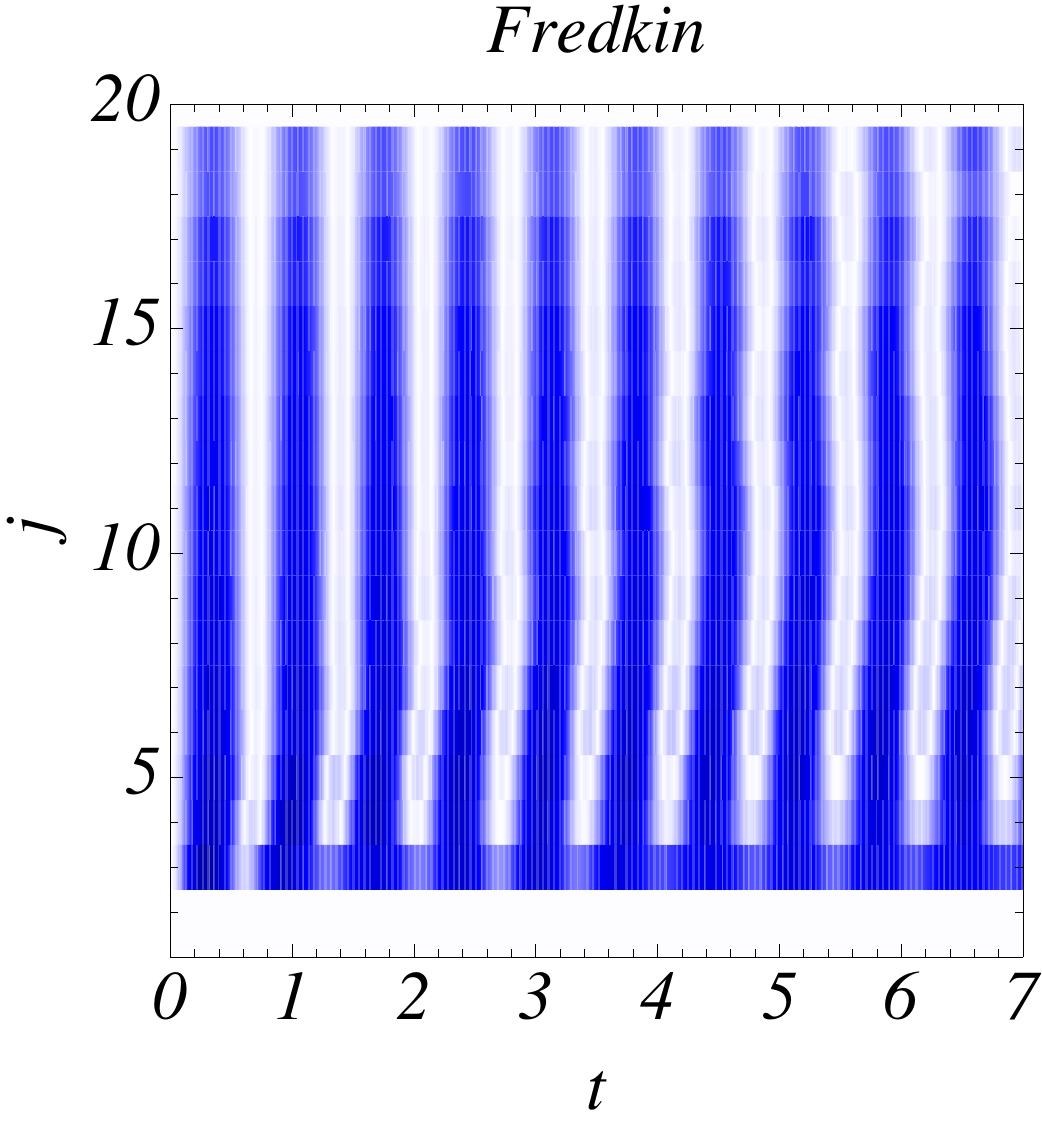}
\caption{
(Left) Time evolution of $\big|\langle S_X(1,t) S_X(j,t)\rangle_c - \langle S_X(1,0) S_X(j,0) \rangle_c\big|$
after switching on, at time $t=0$, the local field $h_z \,S_Z(j_0)$ on the site 
$j_0=1$ of the Motzkin spin chain ($L=20$, $h_z=5$).
(Right) Time evolution of 
$\big|\langle S_X(2,t) S_X(j,t)\rangle_c - \langle S_X(2,0) S_X(j,0) \rangle_c\big|$
after switching on, at time $t=0$, the local field $h_z \,S_Z(j_0)$ on the site 
$j_0=2$ of the Fredkin spin chain ($L=20$, $h_z=5$). The colors are chosen so that white correspond 
to zero and blue of increasing intensity to increasing values. 
}
\label{figure_xx}
\end{figure}

We study, by means of t-DMRG \cite{t_dmrg,note_numerics}, the time evolution of the correlation functions after a local quench. We will consider the time evolution of the $\langle S_Z S_Z\rangle_c$ and $\langle S_X S_X\rangle_c$ correlation functions in both the models, after suddenly switching on a local perturbation $h_zS_Z(j_0)$, 
so that the final Hamiltonia is $H_f=H+h_zS_Z(j_0)$, while the initial state is the ground-state of $H$. We will calculate the time evolution of the excitations after the quench looking at the following quantities
\be
\langle S_Z(i,t)S_Z(j,t)\rangle_c- \langle S_Z(i,0)S_Z(j,0)\rangle_c\equiv \langle e^{iH_f t}S_Z(i)S_Z(j)e^{-iH_f t}\rangle_c -  \langle S_Z(i)S_Z(j)\rangle_c
\ee
the difference between the connected corelation function at $t=0$ and at later time $t$ along the quantization axis, and 
\be
\langle S_X(i,t)S_X(j,t)\rangle_c- \langle S_X(i,0)S_X(j,0)\rangle_c\equiv \langle e^{iH_f t}S_X(i)S_X(j)e^{-iH_f t}\rangle_c -  \langle S_Z(i)S_Z(j)\rangle_c
\ee
the correlators orthogonal to the quantization axis. Notice that the perturbation we choose preserve the $U(1)$ symmetry of the model, so that $\langle S_X\rangle=\langle S_y\rangle=0$ and $\langle S_X(i,t)S_X(j,t)\rangle=\langle S_Y(i,t)S_Y(j,t)\rangle$.\\
As shown in Fig. \ref{figure_zz} for both cases, clear light-cones are visible in the dynamics of $\langle S_Z S_Z\rangle_c$, 
driven by switching-on a local field at one edge of the chain. 
On the other hand the same quench does not produce a light-cone profile in the transverse correlation functions, $\langle S_X S_X\rangle_c$, as one can see from Fig~\ref{figure_xx}. Notice that, for the Fredkin model, we plotted the correlation functions between any points with the second one, since the first site is totally uncorrelated. 

As we can clearly see, the two correlation functions have very different behaviors: $\langle S_ZS_Z\rangle$ behaves as expected by the theory of quantum quenches, namely, the perturbation takes some time in order to reach distant regions, while $\langle S_XS_X\rangle $ shows that the whole system reacts immediately to the perturbation, as being fully causally connected. Because of the arguments reported in the introduction, as we will be seeing in the next section, this anomalous dynamical behavior goes with a violation of the cluster decomposition property for the transverse spin components already at equilibrium. 

\section{Correlation functions}
The anomalous behavior described above can be explained by calculating the correlation functions at equilibrium. 
Since the ground state is a uniform superposition of states which can be mapped to some random walks, we can calculate many ground-state properties resorting to combinatorics \cite{luca16}, see Appendix. In particular we can calculate the magnetization and the correlation functions.  
We remind that the $S_Z$-$S_Z$ connected correlation functions vanish when measured on two points very far apart \cite{luca16}.
Proceeding as in Ref.~\cite{luca16}, one can calculate also the $S_X$-$S_X$ connected correlation functions exactly, showing that, unlike 
  $S_Z$-$S_Z$, they do not vanish even at infinitely large distances, exhibiting 
  therefore a violation of the cluster decomposition property, which is at the origin of
  the unconventional dynamical properties observed numerically and reported in the previous section.  

\subsection{Motzkin model}
\paragraph*{$S_Z$-$S_Z$ correlation functions.}
Defining ${\cal M}^{(n)}_{hh'}$ as the number of Motzkin-like paths
connecting two points at heights $h$ and $h'$ by $n$ steps,
from combinatorics 
the magnetization as a function of the position is given by \cite{luca16}
\begin{equation}
\langle S_Z(i)\rangle
=\frac{1}{{\cal M}^{(L)}}
\sum_{h} {\cal M}_{0 h}^{(i-1)}\left({\cal M}_{h+1, 0}^{(L-i)}-
{\cal M}_{h-1, 0}^{(L-i)}\right)
\label{analitico1}
\end{equation}
where ${\cal M}^{(L)}\equiv {\cal M}^{(L)}_{00}$ is the Motzkin number
(explicit expressions are given in Appendix).
The two-point correlation function 
can be also calculated exactly \cite{luca16} and reads as it follows
\be
\label{SzSz_int}
\langle S_Z(i)S_Z(j)\rangle
=\frac{1}{{\cal M}^{(L)}}
\sum_{h,h'}\,{{\cal M}}^{(i-1)}_{0h} 
\left({\cal M}_{h+1,h'}^{(j-i-1)}-{\cal M}_{h-1,h'}^{(j-i-1)}\right)\left({{\cal M}}^{(L-j)}_{h'+1,0}-{{\cal M}}^{(L-j)}_{h'-1,0}\right)
\ee
The quantity we are interested in is the connected correlation function
\be
\label{szsz_m}
\langle S_Z(i)S_Z(j)\rangle_c=\langle S_Z(i)S_Z(j)\rangle-\langle S_Z(i)\rangle\langle S_Z(j)\rangle
\ee
Examples of this quantity are given in Fig.~\ref{figure_t0} and Fig.~\ref{figure_t0_c} (left plots, blue curves). As shown in those figures, the connected correlation functions go to zero at large distances, for instance
\be
\lim_{L\rightarrow \infty}\langle S_Z(1)S_Z(L)\rangle_c=\lim_{L\rightarrow \infty}\left[
\left(\frac{{\cal M}_{10}^{(L-1)}}{{\cal M}^{(L)}}\right)^2-\frac{{\cal M}_{11}^{(L-2)}}{{\cal M}^{(L)}}\right]=\left(\frac{2}{3}\right)^2-\frac{4}{9}=0
\ee

\paragraph*{$S_X$-$S_X$ correlation functions.}
The transverse correlation functions, $\langle S_X(i)S_X(j)\rangle=\langle S_Y(i)S_Y(j)\rangle$, are given by
\be
\label{sxsx_m}
\langle S_X(i)S_X(j)\rangle
=\frac{1}{{\cal M}^{(L)}}
\sum_{h,h'}\,{{\cal M}}^{(i-1)}_{0h} 
\left({\cal M}_{h-1,h'}^{(j-i-1)}+{\cal M}_{h,h'}^{(j-i-1)}\right)\left({{\cal M}}^{(L-j)}_{h'+1,0}+{{\cal M}}^{(L-j)}_{h',0}\right)
\ee
which is also the connected correlation function, $ \langle S_X(i)S_X(j)\rangle_c=\langle S_X(i)S_X(j)\rangle$ since $\langle S_X(i)\rangle=0$ $\forall i$.
In this case, as shown in Figs.~\ref{figure_t0},  \ref{figure_t0_c} (left plots, red curves), this correlation function does not vanish at infinity distances. For example
\be
\lim_{L\rightarrow \infty}\langle S_X(1)S_X(L)\rangle_c=\lim_{L\rightarrow \infty}\left(
\frac{{\cal M}^{(L-2)}}{{\cal M}^{(L)}}\right)=\frac{1}{9}
\ee
Equation (\ref{sxsx_m}) can be derived introducing 
\be
\label{Spm}
S^\pm=\frac{1}{2}\left(S_X\pm i S_Y \right)
\ee
such that $S^+\Ket{\Uparrow}=0$, $S^+\Ket{0}=\sqrt{2}\Ket{\Uparrow}$ and $S^+\Ket{\Downarrow}=\sqrt{2}\Ket{0}$, and 
$S^-\Ket{\Uparrow}=\sqrt{2}\Ket{0}$, $S^-\Ket{0}=\sqrt{2}\Ket{\Downarrow}$ and $S^-\Ket{\Downarrow}={0}$, and writing 
\be
\label{sxsx_spsm}
\langle S_X(i)S_X(j)\rangle=\frac{1}{4}\Big(\langle S_+(i)S_-(j)\rangle+\langle S_-(i)S_+(j)\rangle\Big)
\ee
since $\langle S_+(i)S_+(j)\rangle=\langle S_-(i)S_-(j)\rangle=0$, because the operators $S^+S^+$ ans $S^-S^-$ project onto finite-total spin sectors, ortogonal to the ground-state. As already pointed out in Ref.~\cite{fradkin}, the quantity $\langle S_+(i)S_-(j)\rangle$ is the probability of finding $\Ket{0}$ or $\Ket{\Downarrow}$ at site $i$ and $\Ket{\Uparrow}$ or $\Ket{0}$ at site $j$, times $\sqrt{2}$. 
This gives half of the expression in Eq.~(\ref{sxsx_m}). Now we can prove that
\be
\langle S_+(i)S_-(j)\rangle=\langle S_-(i)S_+(j)\rangle
\ee
since the action of the operator $S_-(i)S_+(j)$ onto the ground-state is non-zero if, at site $i$ the spin is not $\ket{\Downarrow}$ and at site $j$ the spin is not $\ket{\Uparrow}$, and, in addition, if all the paths beewen the two spins never cross level $1$ (or the horizon $z=1$, according to the definition in Ref. \cite{luca19}). This means that the string of spins between sites $i$ and $j$ should be the Motzkin-like paths with heights lowered by one, namely 
the same paths of those allowed for the $S_+(i)S_-(j)$ operator. As a result, we have simply
\be
\label{sxsx_spsm2}
\langle S_X(i)S_X(j)\rangle=\frac{1}{2}\langle S_+(i)S_-(j)\rangle.
\ee
Our exact result, Eq.~(\ref{sxsx_m}), perfectly agrees with the numerical result for $\langle S_+(i)S_-(j)\rangle$ reported in Ref. \cite{fradkin}, (cfr. Fig.~\ref{figure_t0_c}, left plot) and with DMRG calculation we performed for $\langle S_X(i)S_X(j)\rangle$. We found therefore that, for $L\rightarrow \infty$, $\langle S_X(i)S_X(j)\rangle$ goes to $\frac{4}{9}$ deeply in the bulk and to $\frac{1}{9}$ at the edges.

\subsection{Fredkin model}
\paragraph*{$S_Z$-$S_Z$ correlation functions.}

\begin{figure}[!ht]
\includegraphics[width=8cm]{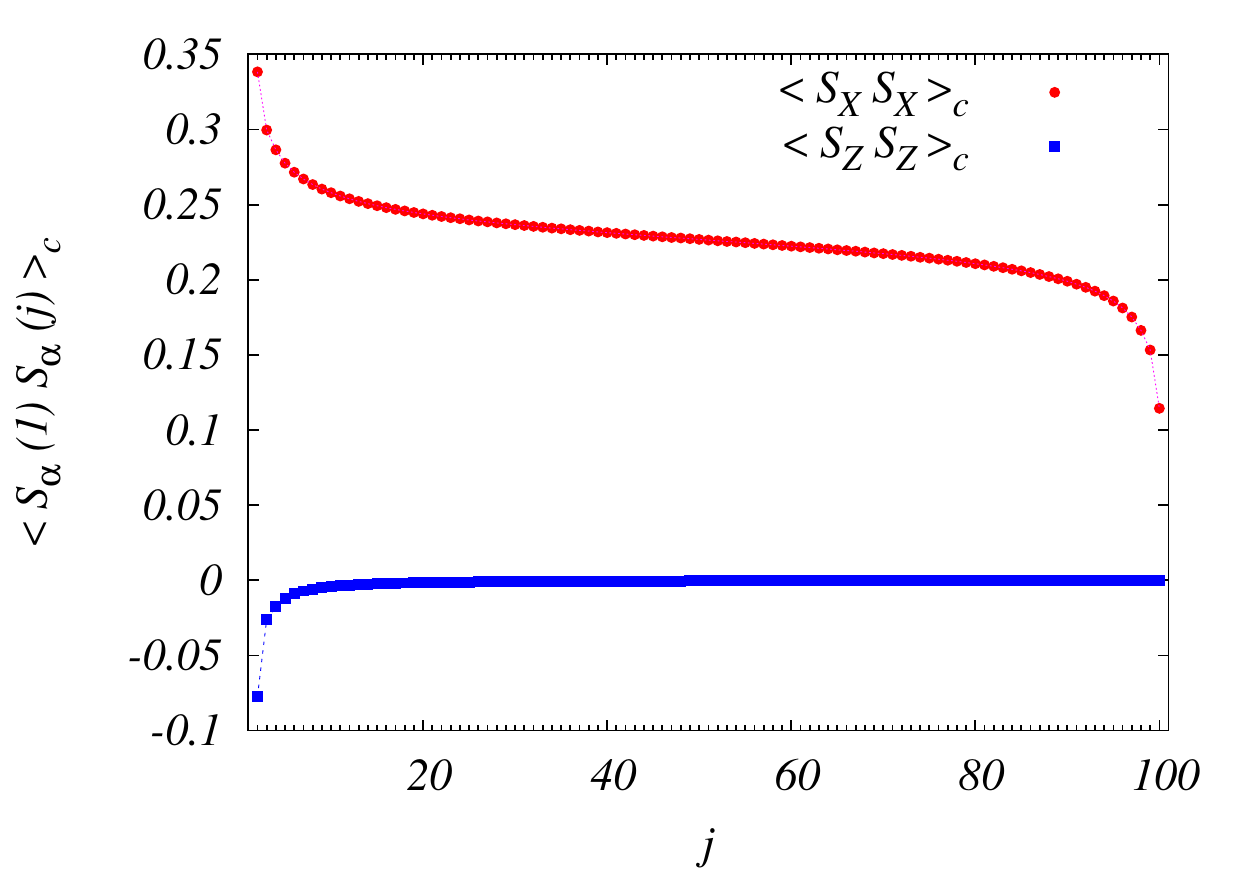}\hspace{1cm}
\includegraphics[width=8cm]{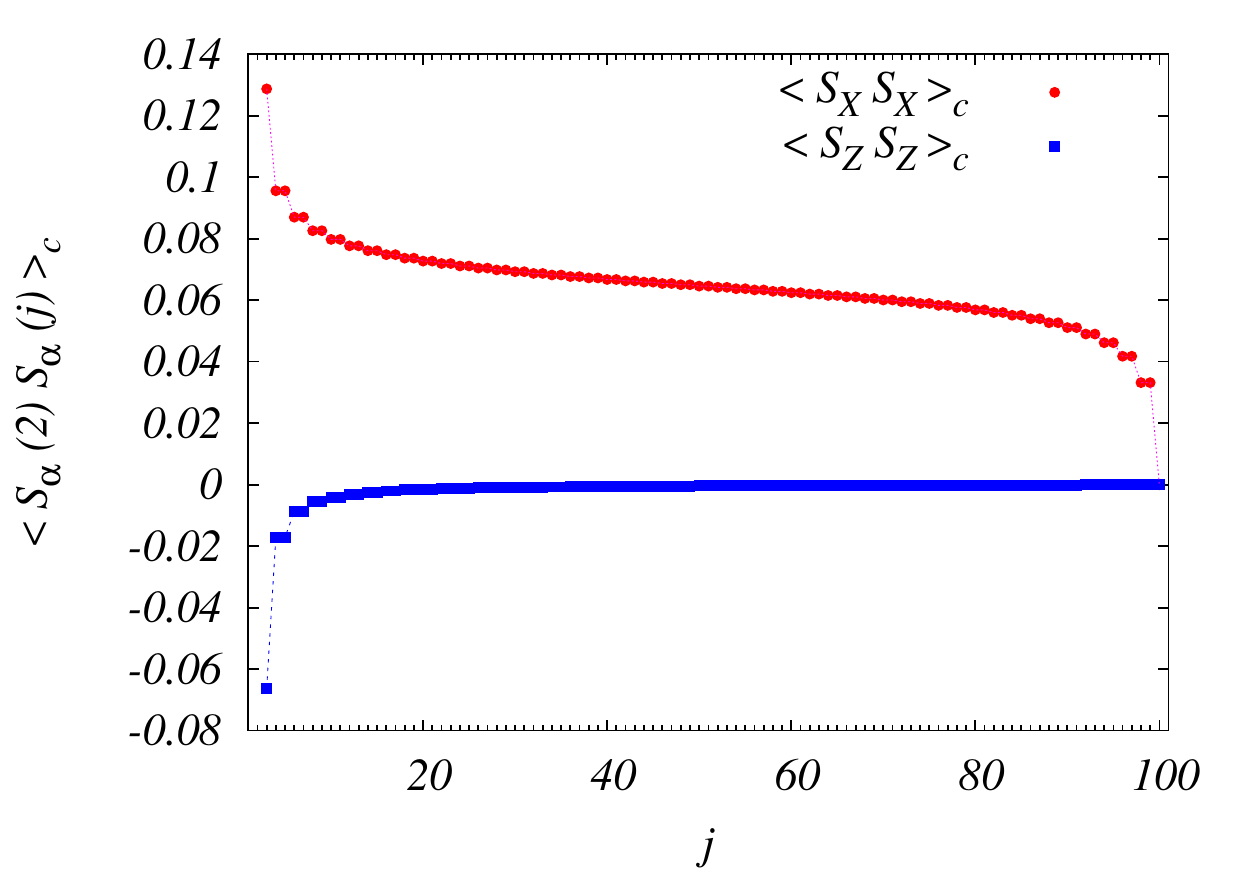}
\caption{(Left) Connected correlation functions  
$\langle S_X(i) S_X(j)\rangle_c$, from Eq.~(\ref{sxsx_m}), and $\langle S_Z(i) S_Z(j)\rangle_c$, from Eq.~(\ref{szsz_m}), with $i=1$, as a function of $j$  
for the Motzkin spin chain ($L=100$). 
(Right) Connected correlation functions  
$\langle S_X(i) S_X(j)\rangle_c$, from Eq.~(\ref{sxsx_f}), and $\langle S_Z(i) S_Z(j)\rangle_c$, from Eq.~(\ref{szsz_f}), with $i=2$, as a function of $j$  
for the Fredkin spin chain ($L=100$).}
\label{figure_t0}
\end{figure}

\begin{figure}[!ht]
\includegraphics[width=8cm]{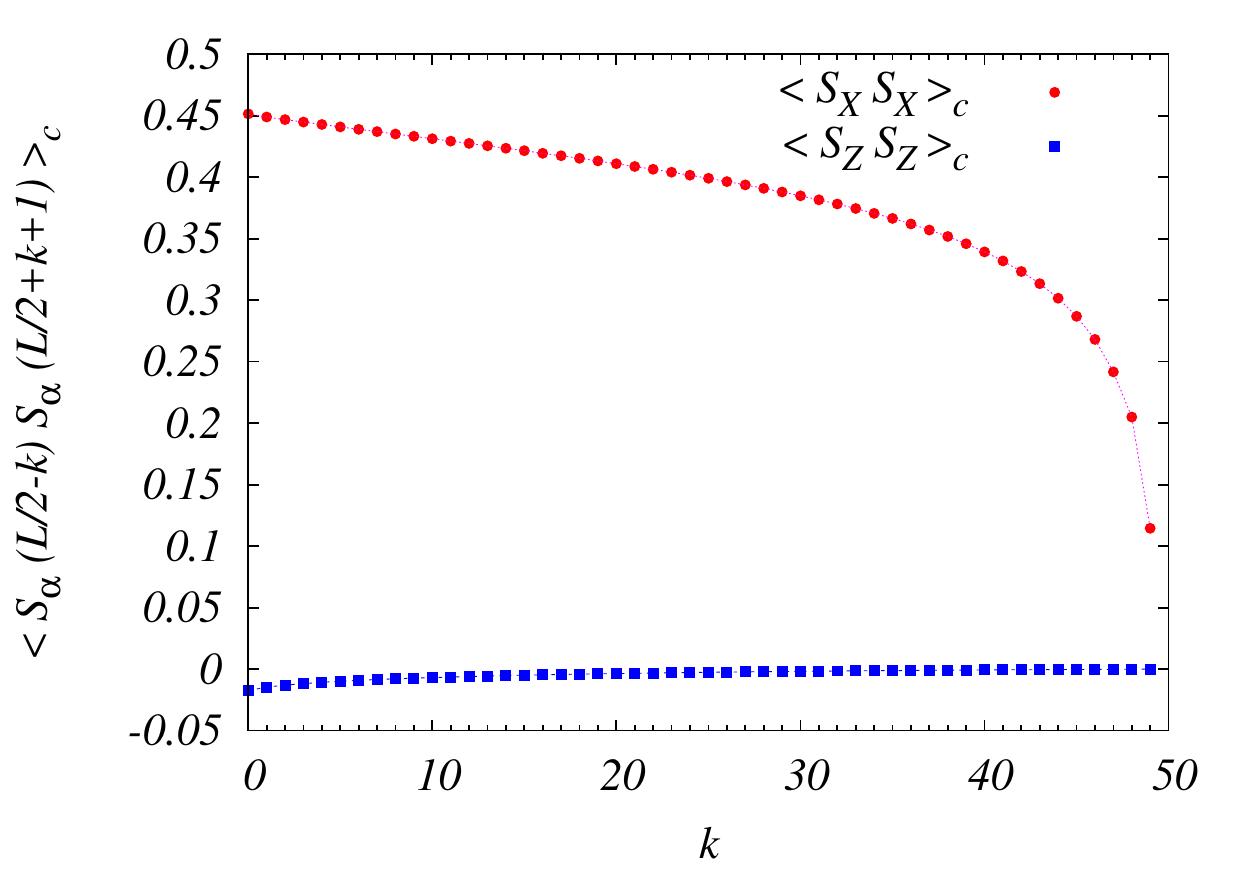}\hspace{1cm}
\includegraphics[width=8cm]{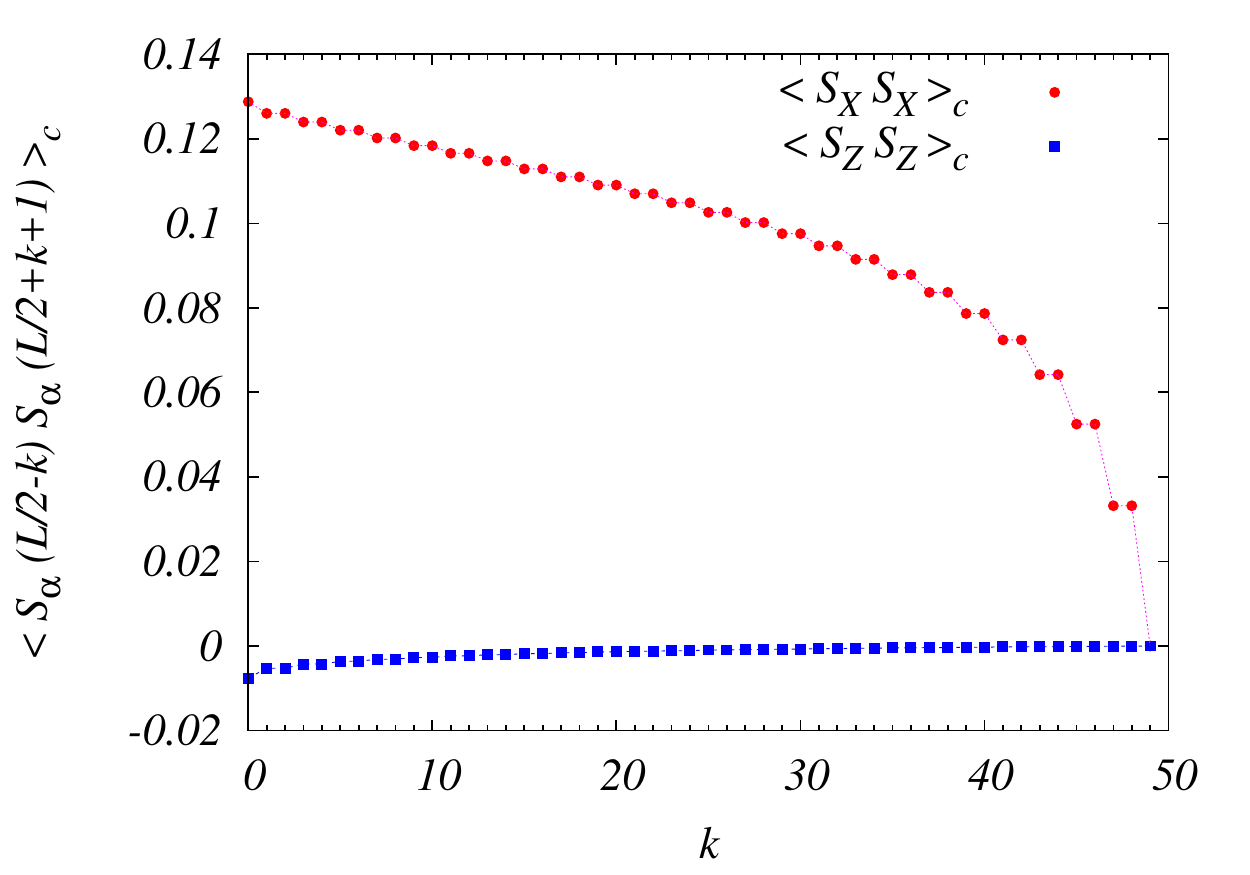}
\caption{(Left) Connected correlation functions  
$\langle S_X(L/2-k) S_X(L/2+k+1)\rangle_c$, from Eq.~(\ref{sxsx_m}), and $\langle S_Z(L/2-k) S_Z(L/2+k+1)\rangle_c$, from Eq.~(\ref{szsz_m}), as a function of $k$  
for the Motzkin spin chain ($L=100$). 
(Right) Connected correlation functions  
$\langle S_X(L/2-k) S_X(L/2+k+1)\rangle_c$, from Eq.~(\ref{sxsx_f}), and $\langle S_Z(L/2-k) S_Z(L/2+k+1)\rangle_c$, from Eq.~(\ref{szsz_f}), as a function of $k$  
for the Fredkin spin chain ($L=100$).}
\label{figure_t0_c}
\end{figure}

We define ${\cal D}^{(n)}_{hh'}$ the number of {\change {paths connecting two points at heights $h$ and $h'$ with $n$ steps, which never cross the ground}}, such that ${\cal D}^{(2n)}\equiv{\cal D}^{(2n)}_{00}=C(n)$, where $C(n)$ are the Catalan numbers (see Appendix for details). 
The magnetizations is, then, given by \cite{luca16}
\begin{equation}
\langle S_Z(i)\rangle
=\frac{1}{2{\cal D}^{(L)}}
\sum_{h} {\cal D}_{0 h}^{(i-1)}\left({\cal D}_{h+1, 0}^{(L-i)}-{\cal D}_{h-1, 0}^{(L-i)}\right).
\label{analitico1_Fredkin}
\end{equation}
%
We can also calculate analytically the correlation functions getting \cite{luca16}
\be
\label{SzSz_hint}
\langle S_Z(i)S_Z(j)\rangle
=\frac{1}{4{\cal D}^{(L)}}
\sum_{h,h'}\,{{\cal D}}^{(i-1)}_{0h}
\left({\cal D}_{h+1,h'}^{(j-i-1)}-{\cal D}_{h-1,h'}^{(j-i-1)}\right)\left({{\cal D}}^{(L-j)}_{h'+1,0}-{{\cal D}}^{(L-j)}_{h'-1,0}\right)
\ee
so that the connected correlation function is, once again, 
\be
\label{szsz_f}
\langle S_Z(i)S_Z(j)\rangle_c=\langle S_Z(i)S_Z(j)\rangle-\langle S_Z(i)\rangle\langle S_Z(j)\rangle
\ee
Some plots of this quantity are reported in Fig.~\ref{figure_t0} and Fig.~\ref{figure_t0_c} (right plots, blue curves). Also in this case, as for the Motzkin model, the $S_Z$-$S_Z$ correlation functions go to zero for large distances, for instance
\bea
\nonumber
\lim_{L\rightarrow \infty}\langle S_Z(2)S_Z(L-1)\rangle_c&=&\frac{1}{4}\lim_{L\rightarrow \infty}\left[
\left(\frac{{\cal D}_{20}^{(L-2)}-{\cal D}^{(L-2)}}{{\cal D}^{(L)}}\right)^2-\frac{{\cal D}^{(L-4)}_{22}-2{\cal D}^{(L-4)}_{02}+{\cal D}^{(L-4)}}{{\cal D}^{(L)}}
\right]\\
&=&-\lim_{L\rightarrow \infty}\frac{3(L+2)}{8(L-1)^2(L-3)}=0
\eea

\paragraph*{$S_X$-$S_X$ correlation functions.}
The transverse correlation functions, $\langle S_X(i)S_X(j)\rangle=\langle S_Y(i)S_Y(j)\rangle$, are given by
\be
\label{sxsx_f}
\langle S_X(i)S_X(j)\rangle
=\frac{1}{2{\cal D}^{(L)}}
\sum_{h,h'}\,{{\cal D}}^{(i-1)}_{0h} {\cal D}_{h-1,h'-1}^{(j-i-1)} {{\cal D}}^{(L-j)}_{h',0}
\ee
which is also the connected correlation function, $ \langle S_X(i)S_X(j)\rangle_c=\langle S_X(i)S_X(j)\rangle$ since $\langle S_X(i)\rangle=0$ $\forall i$. As shown in Figs.~\ref{figure_t0},  \ref{figure_t0_c} (right plots, red curves), it does not vanishes at infinity distances. For instance, 
\be
\lim_{L\rightarrow \infty}\langle S_X(2)S_X(L-1)\rangle_c=\frac{1}{2}\lim_{L\rightarrow \infty}\frac{{\cal D}^{(L-4)}}{{\cal D}^{(L)}}=
\lim_{L\rightarrow \infty}\frac{L(L+2)}{32(L^2-4L+3)}=\frac{1}{32}
\ee
Notice that the first and the last spin are completely uncorrelated with the rest of the chain, so that we considered the second and second-last spins. 
In order to derive Eq.~(\ref{sxsx_f}), one can introduce the operators as in Eq.~(\ref{Spm}), 
such that $S^+\Ket{\uparrow}=0$, $S^+\Ket{\downarrow}=\Ket{\uparrow}$, and 
$S^-\Ket{\downarrow}={0}$, $S^-\Ket{\uparrow}=\Ket{\downarrow}$. 
One can write $\langle S_X(i)S_X(j)\rangle$ as in Eq.~(\ref{sxsx_spsm}) and realize that $\langle S^+(i)S^-(j)\rangle$ is the probability $P$ of finding spin down at site $i$ and spin up at site $j$. Repeating the argument described before one gets that $\langle S^-(i)S^+(j)\rangle=\langle S^+(i)S^-(j)\rangle$ so that 
Eq.~(\ref{sxsx_spsm2}) is valid. One obtain, therefore, Eq.~(\ref{sxsx_f}) which is half of $P$, the probability of having spin down and spin up at $i$ and $j$ respectively. 
Also in this case our exact result, Eq.~(\ref{sxsx_f}), is in perfect agreement with the numerical result for $\langle S_+(i)S_-(j)\rangle$ reported in Ref. \cite{fradkin}, (cfr. Fig.~\ref{figure_t0_c}, right plot) and with DMRG calculation we performed for $\langle S_X(i)S_X(j)\rangle$. We find that, for $L\rightarrow \infty$, $\langle S_X(i)S_X(j)\rangle$ goes to $\frac{1}{8}$ deeply inside the bulk and to $\frac{1}{32}$ at the edges, for second and second-last spins.

\section{Conclusions}
We performed a quantum quench by applying a perturbation which preserve the $U(1)$ symmetry of the model and observe a light-cone propagation of the excitations looking at the correlation functions of spins along $z$-directions, consistently with the vanishing behavior of the connected correlation functions at equilibrium. On the contrary, looking at the correlation functions of spins along the transverse directions we do not observe a light-cone because the system is already long-range correlated, as shown by exact results for the $S_X$-$S_X$ correlation functions. This work extends significantly the analysis done in a previous paper \cite{luca16}, where violation of the cluster decomposition has been observed for the colorful versions of the models looking at the correlation functions of spins along $z$-directions. The results reported here are in perfect agreement with numerical results for transverse spin correlation functions \cite{fradkin} and with what found for the mutual information, which remains finite even when measured between infinitely distant separated regions \cite{luca19}. 
In this paper we considered the colorless versions of the Motzkin and Fredkin spin chains, with spins $1$ and $1/2$. 
It would be very interesting to study the transverse spin correlations and the related dynamics also in the corresponding colorful models, namely with higher values of the spins. 
  
\begin{acknowledgments}
\noindent We thank X. Chen, E. Fradkin, I. Klich, V. Korepin, O. Salberger, W. Witczak-Krempe for useful discussions.  L.D. thanks also SISSA for kind hospitality. L.B. acknowledges ERC Starting Grant TopoCold for financial support.  
\end{acknowledgments}

\appendix
\section{Combinatorics}
Let us define $p_n=(1-\textrm{mod}(n,2))$ such that $p_{2n+1}=0$ and $p_{2n}=1$, 
so that it selects only even integer numbers, and
\be
{\cal D}^{(n)}_{hh'}=
\left[\left(\ba{c}
n\\\frac{n+|h-h'|}{2}\ea\right)-
\left(\ba{c}
n\\\frac{n+h+h'}{2}+1\ea\right)\right]p_{n+h+h'}
\ee
where ${\cal D}^{(n)}_{hh'}$ are the number of Dyck-like paths between two points at distance 
$n$ and heights $h$ and $h'$. In particular
\bea
{\cal D}^{(n)}\equiv {\cal D}^{(n)}_{00}=C\left({n}/{2}\right)\,p_n
\eea
with 
\be
C(\ell)=\frac{(2\ell) !}{\ell !(\ell+1)!}
\ee
the Catalan numbers. Useful relations are
${\cal D}^{(n)}_{h0}=\frac{h+1}{\frac{n+h}{2}+1}
\left(                               
\ba{c}                                                                         
n\\                                                                            
\frac{n+h}{2}                                                                  
\ea                                                                            
\right)p_{n+h}$
and ${\cal D}^{(n)}_{0h}={\cal D}^{(n)}_{h0}$. Let us also define 
\be
{\cal M}^{(n)}_{h h'}=\sum_{\ell=0}^{\lf\frac{n-|h'-h|}{2}\rf}
\left(\ba{c} n\\2\ell+|h'-h|\ea\right)
{\cal D}^{(2\ell+|h'-h|)}_{hh'}
\ee
the number of Motzkin-like paths between two points at heights 
$h$ and $h'$ connected by $n$ steps. In particular
\be
{\cal M}^{(n)}\equiv {\cal M}^{(n)}_{00}=\sum_{\ell=0}^{\lfloor\frac{n}{2}\rfloor}
\left(\ba{c} n\\
2\ell\ea\right)
C(\ell)
\ee
are called Motzkin numbers.




\begin{thebibliography}{99}
\bibitem{note_numerics} In all our dynamical simulations the ground state is found by keeping up to 1400 DMRG states and performing 5 finite size sweeps. The dynamics is computed by using up to 1024 DMRG states and a time step $\delta t=0.01$.
\end{thebibliography}
\end{document}